\documentclass[nocite]{epl}

\title{Two-dimensional  shear modulus of a Langmuir foam}
\shorttitle{2D shear modulus of a Langmuir foam}
\author{
S. Courty\inst{1}  \protect\cite{present}, B. Dollet\inst{1} ,
F. Elias\inst{2} \protect\cite{federation}, 
P. Heinig\inst{3} 
\and
F. Graner\inst{1}  
\thanks{ $\;$
Author for correspondence at
graner@ujf-grenoble.fr. Fax: (+33) 4 76 63 54 95.}
}
\shortauthor{S. Courty \etal}
\institute{\inst{1}
 Laboratoire de Spectrom\'etrie Physique (UMR 5588 CNRS -
Universit\'{e} J. Fourier Grenoble 1), BP 87, F-38402 Saint Martin
d'H\`eres cedex,  France
\\
  \inst{2}
Laboratoire des Milieux D\'esordonn\'es et
   H\'et\'erog\`enes (UMR 7603 CNRS - Universit\'e Paris
6), case 78, 4 place Jussieu, F-75252 Paris cedex 05, France
\\
\inst{3} 
Max Planck Institut of Colloids and Interfaces, Am M\"uhlenberg 1, D-14476 Golm, Germany
}

\begin{document}

\maketitle

%\date{\today}

\begin{abstract}
We deform a two-dimensional (2D) foam, created in a Langmuir monolayer,  
by applying a mechanical perturbation, and simultaneously
image it by Brewster angle microscopy.  We determine the foam stress tensor (through a 
determination of the 2D gas-liquid  line tension, 2.35 $\pm$ 0.4 
pJ$\cdot$m$^{-1}$)
and the statistical strain 
tensor, by analyzing the images of 
the deformed structure. We deduce the 2D shear modulus of the foam, $\mu= 38 \pm 3\; 
\mathrm{nN}\cdot \mathrm{m}^{-1}$.
 The foam 
effective rigidity is predicted to be $  35 \pm 3\; \mathrm  {nN}\cdot \mathrm  {m}^{-
1}$, which agrees with  the value  $37.5 \pm 0.8\; \mathrm {nN}\cdot \mathrm 
{m}^{-1}$ obtained in an independent 
mechanical measurement.

\end{abstract}

\bigskip

\section{Introduction}

A liquid foam, made of polyhedral gas bubbles separated by 
thin liquid 
walls forming a connected network  \cite{weaire_bible}, is a mixture of two fluids.  It has 
nevertheless a solid-like  elasticity, characterised by a 
shear modulus  
$\mu$,  proportional to the surface tension of the walls
\cite{hohler,cohenaddad}.
% (and independent  on the pressure  of the bubbles, 
%which does not oppose shear). 
In fact, shearing a foam modifies the total 
length of the walls, thus  the foam energy.  
The value of $\mu$ can be determined in numerical simulations
 \cite{weaire,kawasaki,durian}; however,   it is still an open problem to predict analytically 
its value for 
a real foam,  which has a finite fluid fraction and an inherent disorder due to its distribution 
of  bubble sizes.

Here, we compare two
experimental measurements of $\mu$.
First, 
by  global
mechanical measurements on the scale of  the 
whole  foam, described in terms of elasticity of continuous media.  
 Second, and simultaneously, 
by detailed imaging  of the diphasic foam structure, on the 
local level of a few bubbles: this
 suggests to use two-dimensional (2D) foams. In the 
literature, 2D soap froths have been sheared 
in Couette geometry, either as bubble rafts 
\cite{lauridsen, abdelkader,pratt}  or  confined  in Hele-Shaw cells between two parallel plates of 
glass  
\cite{debregeas}. 

We  investigate the elasticity of a real  2D system:
a  ``Langmuir foam''  \cite{losche}.  A monomolecular layer 
of amphiphilic molecules deposited at the surface of water (``Langmuir 
monolayer")  
exhibits a first 
order transition between a 2D gas phase and a denser 2D liquid
 (also called  ``liquid-expanded") phase.  In the 2D gas-liquid coexistence region, the 
domains 
spontaneously arrange into a foam \cite{losche}. Walls are stabilised by electrostatic dipolar 
interactions 
of the molecule themselves, without any added external surfactant 
\cite{mann,riviere}. 
Such Langmuir foams are approximately characterised by a line tension (see below), 
hence obey the same Plateau rules as other 2D liquid foams
\cite{stine, berge}, and display the same generic rheological 
behavior \cite{dennin}.  

At  the global level, we probe the foam effective rigidity  $k_{\mathrm{eff}}$  
by 
measuring  the force exerted on 
a rigid obstacle moving relatively to the foam 
\cite{amorphes}. Independently, at the local level, we determine the foam stress tensor 
(through a determination of the 2D gas-liquid  line tension)
and the statistical strain 
tensor, by analyzing the 
 bubble deformation   \cite{aubouy}; we deduce the 2D shear modulus of the 
foam 
\cite{asipauskas}. We then compare both 
measurements,
 in the 
frame of linear elasticity of continuous isotropic media \cite{landau}.

\section{Methods}

  %\subsection{preparation du systeme  }

The 2D Langmuir foam is formed at $T= 21^\circ$C
in a home-made Langmuir teflon (PTFE) trough, of dimension 
$a\times b = 8 \times 2$~cm$^2$, 0.7 cm deep.  
 Pentadecanoic acid (C$_{14}$H$_{29}$CO$_{2}$H)  is dissolved in 
chloroform at a concentration of $3 \cdot 10^{-4}$ mol$\cdot$l$^{-1}$.
In order to prevent 
the dissociation of acid,  the pH of the  ultrapure 
water is set at pH=2 by adding hydrochloric acid  \cite{stine, berge}.  
The 
air-water interface is cleaned by aspiration, then $\sim 10$ $\mu$l
of the solution are spread onto the surface of water.  After 10 
minutes, the chloroform is evaporated, and a  
Langmuir foam of pentadecanoic acid forms.   Using teflon compression barriers, 
we choose to adjust the surface fraction of 2D liquid  to  the limit  at which walls are robust and clearly visible, which is about  23~\%.
With an average bubble size 
$\sim$100~$\mu$m, bubbles are large enough for image analysis, and there are enough 
bubbles to 
perform statistics.

%\begin{figure}
%\begin{center}
%\onefigure[width=6cm]{Fig_setup.eps}
%%\epsfig{file=FigBam.eps, width=3.1in}
%\end{center}
%\caption{
%Experimental set-up.   The Langmuir foam, created at the  surface of water, 
%is sheared by the free end of a soft optical fiber  displaced in the plane of the monolayer.  
%It  is visualised using a Brewster angle  microscope (BAM): laser 1 and camera (as well as 
%polarisers, not shown). Simultaneously,  the foam resistance, is measured by detecting the 
%deflection  of the  laser beam (laser 2) transmitted by the fiber under the trough.   }
%\label{Fig set_up}
%\end{figure}
%

%\subsection{mecanique}

The experimental set-up,  
%(Fig.  \ref{Fig set_up}),
inspired from ref. \cite{barentin1},   is 
presented in ref. 
\cite{amorphes}. We determine
the resistance exerted by the foam on an obstacle displaced relatively to the 
foam  in the horizontal plane.
The obstacle is the tip of a vertical rod, actually a denuded glass optical 
fiber (Thorlabs), which also acts as force sensor. 
Since the shear modulus of the Langmuir foam is low, we must here take 
a much softer fiber than in  ref. 
 \cite{amorphes}.
We chose a length 
$L=3.5\pm 0.1$ cm, then attack the glass fiber with fluorhydric acid 
at 40\% concentration until we reach its core, of diameter 
$2r = 6\pm 1$~$\mu$m.

\begin{figure}
\begin{center}
\onefigure[width=14cm]{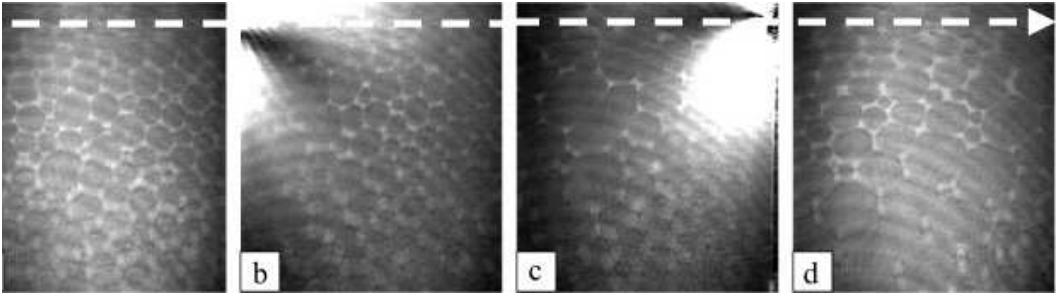}
\end{center}
\caption{Four images of the movie \protect\cite{film_web}:
(a) 10.5 s before, (b) 0.7 s before, (c) 
0.7 s after and (d)  
28.4 s after the fiber passes through the middle of the field of view.
The fiber moves from the left 
to the right (white arrow)  at a constant applied velocity $\dot{X}_{\mathrm{imp}} =620$ 
$\mu$m$\cdot$s$^{-1}$.
The size of each 
image
(here corrected by the projection factor at Brewster angle)  is 768 $\times$ 
850 $\mu$m$^2$.
The 2D gas phase of the monolayer, of low density, 
corresponds to a zero reflected intensity and appears black on the 
images, whereas the liquid phase, denser, appears brighter \protect\cite{henon}. 
The average area is larger on the last images than on the first ones: this is due to the large-scale inhomogeneities (unavoidable with our preparation method) made apparent by the bulk flow, and not to the increase of area of each bubble separately
(which contributes less than 6 \% during  our experiment, and is likely due to a partial solubilization of amphiphilic molecules in water, rather than to actual coarsening \cite{heinig}). We do not observe wall breakage (it appears only at 4 times higher velocity, or with a stiffer fiber), nor perturbation 
of the liquid-gas coexistence.
 }
\label{Fig_images}
\end{figure}

The lower   end of the fiber   plunges vertically in the foam 
and immerses 10 
$\mu$m below the surface.  The fiber is silanised 
(n-octadecyltrichlorosilane diluted at 2\% in octane):  the
contact angle, measured
with a camera attached to the side of the trough, is 
 close to 90$^\circ$, and the  residual  meniscus is small  (white 
region on Fig.  
\ref{Fig_images}b, c).

The 
upper end of the fiber  is held in a 
concentric chuck 
fixed on a horizontal translation stage coupled to a 
motor, allowing for a horizontal displacement $X_{\mathrm{imp}}$:
this  
applies a horizontal point-like deformation 
to the monolayer.  
The horizontal deflection $\zeta$ of the free end of the fiber  
%(Fig.  \ref{Fig set_up})
%with respect to the displacement of its fixed end,
 is then measured by 
connecting the fiber to a laser diode, and collecting (in a 
photodetector placed under the trough)  the beam transmitted at the free 
end of the fiber. We then deduce the horizontal resistance force $F$ exerted by 
the 
foam on the vertical fiber as $F=K\zeta$. 
Here, the fiber rigidity $K$ 
(calibrated by holding the fiber horizontally at one end, and 
measuring its deflection under its own weight) is $K = 362\pm 
30$ nN$\cdot$m$^{-1}$, suitable for precise measurements in the pN range.

The foam effective rigidity 
$k_{\mathrm{eff}}$, {\it i.e.} 
the ratio of the force $F$ to the horizontal displacement $X = 
X_{\mathrm{imp}}-\zeta$ of the fiber within the foam, is measured as
$k_{\mathrm{eff}} = F/X = K\zeta/X$.
%  (Fig.  \ref{Fig set_up}).
It characterises the whole foam  (it depends on its elastic moduli, as well as its size and 
geometry) under a given applied deformation, as long as it remains   elastic.
It is physically intuitive: $k_{\mathrm{eff}}$  represents the rigidity  one would feel by 
sticking a finger 
in the foam and moving it laterally.

  %\subsection{image}

The contrast
and resolution of the Brewster angle microscope \cite{henon} 
% (Fig.  \ref{Fig set_up}) 
have been optimised to image this foam;  for  details
see ref.  \cite{courtythese} pp. 54-60.  Briefly, an 
incident He-Ne (Uniphase, 30 mW, 632.88 nm) laser beam is tilted at an 
incident angle equal to the Brewster angle of the air-water interface 
$i_{\mathrm{B}}  \approx 53^\circ$, passes a Glan Thomson
polariser (Melles-Griot) in the plane of incidence, and a quarter wave plate
(Melles-Griot).
After 
reflexion on the surface of water, the beam 
 enters a combination of two objectives (Zeiss) with small numerical 
aperture (NA = $0.1$), magnifying 5 and 10 times, respectively,
and an analyser (Newport).  The 
image of the surface forms on a CCD camera (sensitivity 10$^{-3}$ lux) 
and is recorded simultaneously on video and on computer {\it via} a 
Scion Image frame grabber.  

  %\subsection{image analysis}

On each image (Fig.   \ref{Fig_images}),
 we measure the contribution of the wall  network to the stress tensor, as follows  
\cite{reinelt}. 
In a 2D foam,  a wall represents two gas-liquid interfaces: thus
the wall tension is $\tau  =2\lambda$, where $\lambda$ is the gas-liquid line tension (see 
below). 
The network contribution to the stress   $\overline{\overline{\sigma }} $
could  be  measured from the image,  by identifying the bubble walls which cross a given line 
of  unit length and normal vector $\vec{v}$.
The vectorial sum of their tensions \cite{landau}, here 
 $\vec{\tau}= 2\lambda 
\hat{e}$, where $\hat{e}$ is the 
unit vector tangent to the 
wall (its orientation is unimportant in what follows), would measure
$\overline{\overline{\sigma }}\cdot \vec{v} $.
As shown in ref. \cite{aubouy}, we measure the stress tensor with better statistics
\cite{asipauskas}
if we use its equivalent definition over the  bulk of the image
\cite{kraynik}, which writes here (neglecting the curvature 
of walls 
\cite{courbure}): 
  $\overline{\overline{\sigma }} = S^{-1} \sum 
\vec{\tau} \otimes \vec{\ell}$, 
where $S$ is the area of the image, $\vec{\ell} = \ell \hat{e}$; $\ell$ is the 
length of the wall; the sum is taken over all the walls present on the image; 
$\otimes$ is the tensor product:
$\sigma_{ij} =   S^{-1} \sum  \tau _{i} \ell_{j}  $ 
\cite{aubouy}.  

\begin{figure}
\begin{center}
%\onefigure[width=13.5cm]{Fig5-15(b).eps}
%\onefigure[width=14cm]{Fig_setup_lambda.eps}
\onefigure[width=14cm]{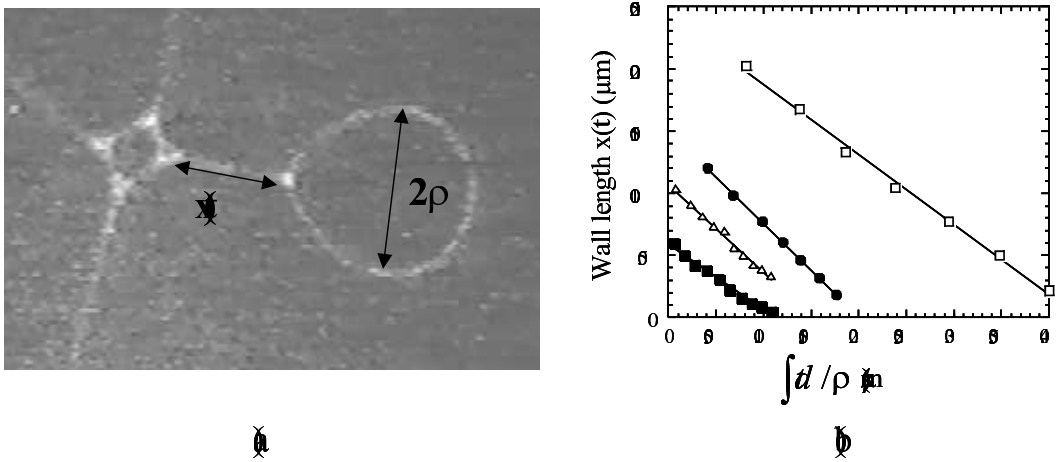}
\end{center}
\caption{ Measurement of the line tension $\lambda$.
(a) Image (320 $\times$ 215 $\mu$m$^2$) by fluorescence microscopy. By breaking a wall (on 
the right of the image, not 
shown), we obtain a bubble (radius $\rho$), attached to a single wall (length $x$), which 
retracts due to the line tension.
(b) Plot of wall length $x(t)$ {\it versus} $\int_0^t dt / \rho(t)$ (see text) for four 
experiments with different initial length $x(0)$. 
We obtain straight lines,  which proves that, at these length
scales, the line tension is constant despite long-range dipolar interactions
\protect\cite{mann,riviere,heinig,fischer_friction}. 
}
\label{Fig_relax}
\end{figure}

We thus  need to determine the gas-liquid line tension $\lambda$  (Fig. 
\ref{Fig_relax}a).
We adapt the method of ref.  \cite{fischer_break}, as follows.
We record images  in fluorescence microscopy, with 1\% NBD-HDA dye (all other 
conditions being unchanged). With a localised laser heating (2 K), we break one wall.
The resorption of  the broken wall's free extremity is driven by the wall 
tension $\tau=2\lambda$. It works 
 against the dissipation force, $F_{\mathrm{drag}}$.
Since the monolayer is not in a dense phase, 
the usual 3D viscosity   $\eta_{\mathrm{3D}} $ of the water subphase dominates the 2D one
\cite{klinger,fischer3,fischer4}, and: 
$ F_{\mathrm{drag}} = - (3\pi^2/4) \; \rho\;  \eta_{\mathrm{3D} }\; \dot{x}$.
Here  $\dot{x}$ is the rate of 
retraction of the length $x$ of the wall; $\rho$ is the radius of the object at the end 
of the wall. We chose the largest possible $\rho$ (to enhance sensitivity): here, a free bubble 
remains attached at the end of the wall (Fig. \ref{Fig_relax}a). The 
prefactor $(3\pi^2/4) \approx 7.4$ is calculated at the limit of vanishing 
surface viscosity (zero Boussinesq number) for a deformable object (here: the 
bubble)  \cite{deKoker}, as opposed to a solid object which would lead to a prefactor of  8  
\cite{fischer_friction,fischer4}. 
We then deduce $\lambda$ from the balance of both forces. Since the bubble radius $\rho(t)$ 
can vary with time, we 
write:
$x(t)-x(0) = - 8 \lambda (3 \pi^2 \eta_{\mathrm{3D}})^{-1} \int_0^t dt / \rho(t).$
From the  slopes of Fig. (\ref{Fig_relax}b),  we measure  $\lambda$ = 2.35 $\pm$ 0.4 
pJ$\cdot$m$^{-1}$.

To quantify the bubble anisotropy, we use  
the local {\it texture tensor} $\overline{\overline{M}}$ \cite{aubouy}. It is
a tensor constructed using all walls  in a local region (here: the 
field of view)
of the foam:
$\overline{\overline{M}} = \langle \vec{\ell}\otimes \vec{\ell}\rangle 
$, {\it i.e.} $M_{ij} = \langle \ell _{i} \ell _{j}\rangle$, where
$\vec{\ell}$ is the vector linking both ends of a wall,
and $\langle \rangle$ stands for an 
average over the walls \cite{aubouy}.
Its logarithm $\ln \overline{\overline{M}}$ 
has the same 
axes as $\overline{\overline{M}}$, and is real and symmetric
\cite{log_matrice}. We use it to define the {\it statistical strain tensor} 
$\overline{\overline{U}}
= (\ln \overline{\overline{M}}-\ln 
\overline{\overline{M}}_{0})/2$, where 
$\overline{\overline{M}}_{0}
%=\langle \ell _{0}^{2}\rangle\overline{\overline{I}}/2
$ is the reference value of 
$\overline{\overline{M}}$ in the undeformed state of the foam.
%, $\ell_{0}$ being the average wall length \cite{aubouy,asipauskas}.  
This tensor   $\overline{\overline{U}}$  
reduces to the usual definition of strain in the validity limits of 
classical elasticity \cite{aubouy}.

\begin{figure}
\begin{center}
%\twofigures[scale=0.5]{Fig5-21.eps}{Fig5-22bis.eps}
\onefigure[width=14cm]{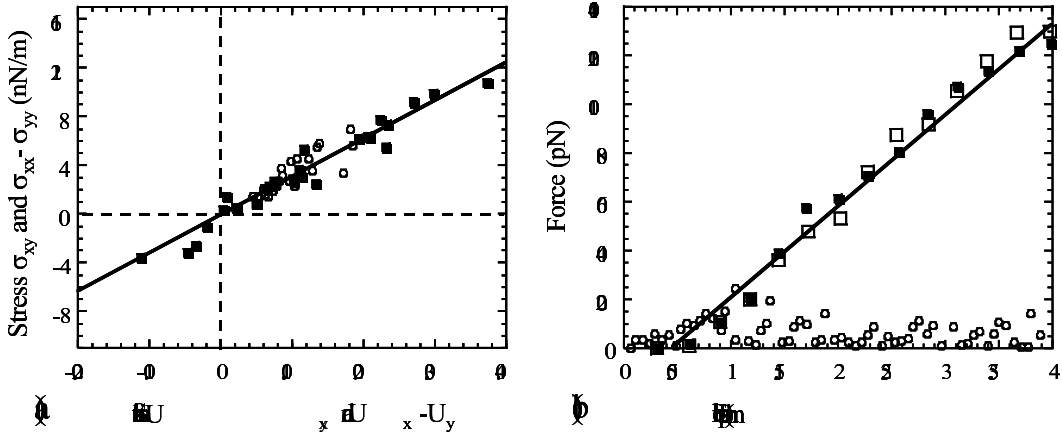}
\end{center}
\caption{Two independent determinations of the Langmuir foam elasticity.
(a) Measurement of the shear modulus $\mu$ based on analysis of 28 images.  Each 
image provides two points, one for each component:
closed squares: normal stress 
difference $\sigma_{xx}-\sigma_{yy}$ {\it versus} 
$U_{xx}-U_{yy}$; open 
circles: shear stress $\sigma_{xy}$ {\it versus} 
$U_{xy}$. Solid line: linear fit through all points, slope 
$76\pm 6$~nN$\cdot$m$^{-1}$.  
(b) Mechanical measurement of the effective rigidity $k_{\mathrm{eff}}$. 
The
force $F =  K \zeta $ exerted by the foam 
on the fiber is plotted {\it versus} the displacement $X
= X_{\mathrm{imp}} - \zeta$
of the fiber free end. 
Closed squares:
the fiber is displaced in the direction $X_{\mathrm{imp}}>0$ 
at a constant velocity of 620 $\mu$m$\cdot$s$^{-1}$. Open 
squares: it 
returns back to its initial position at the same velocity.   Solid line: linear fit 
through all points, slope 
$37.5\pm 0.8$~nN$\cdot$m$^{-1}$.  
 Open circles: control experiment, at the surface of pure water.
}
%\label{Fig courbe_image}
\label{Fig_courbes}
\end{figure}

We measure the two components $U_{xy}$ and $U_{xx}-U_{yy}$:
they are independent of $\overline{\overline{M}}_{0}$; in a weakly deformed material, 
they
are roughly equal to $ \langle \ell_x^2 -\ell_y^2  \rangle/ \langle 
\ell_x^2 + \ell_y^2  \rangle$ 
and $\langle \ell_x \ell_y  \rangle/ \langle \ell_x^2 + \ell_y^2  \rangle$, 
respectively.
We then compare them to the corresponding components of the stress tensor 
(Fig.  \ref{Fig_courbes}a)  \cite{asipauskas}:
the relation between the stress and
strain tensors  defines   the shear modulus $\mu$ \cite{landau}.
In fact, in a 2D linear, homogeneous and isotropic medium, where the stress has no 
vertical diagonal 
component  (``plane stress", $\sigma _{zz} =0$), the 2D Hooke law  defines $\mu$,
and the 2D Poisson ratio $\nu$ ($-1 \le  \nu  \le 1$), through
\cite{amorphes}:
\begin{equation}
\sigma _{xx}+\sigma _{yy} =
2\mu\; \frac{1+\nu }{1-\nu }\; (U_{xx}+U_{yy}), \quad
\sigma _{xx}-\sigma _{yy} =
2\mu \; (U_{xx}-U_{yy}), \quad
\sigma _{xy} =
2\mu U_{xy}.
\label{eq_elasticity}
\end{equation}

\section{Results}

%\subsection{courbe image}

Initially (Fig. \ref{Fig_images}a),  the bubbles are relaxed and the foam is isotropic.  
The fiber approaches the field of view  (Fig. \ref{Fig_images}b) and compresses the bubbles.  
 Behind the fiber  (Fig. \ref{Fig_images}c), the bubbles are stretched.
 After relaxation (Fig. \ref{Fig_images}d) the foam becomes again isotropic.
Fig.  (\ref{Fig_courbes}a) shows a plot of the elastic normal stress 
difference $\sigma_{xx}-\sigma_{yy}$ {\it versus} $U_{xx}-U_{yy}$, and 
of the shear stress $\sigma_{xy}$ {\it versus} $U_{xy}$.  All the data 
collapse on the same straight line;   its slope measures $2\mu$ 
(eq. \ref{eq_elasticity}):
$$\mu= 38 \pm 3\; \mathrm{nN}\cdot \mathrm{m}^{-1}.$$

 This value 
can be compared to the   theoretical computation  for a 2D foam with a regular, dry honeycomb 
structure \cite{kraynik,princen,morse} of bubble area $A$: 
$\mu_{\mathrm{h}} = 0.465 \times (2\lambda) 
A^{-1/2}$. 
For our system, $2\lambda = 4.7$ pJ$\cdot$m$^{-1}$, the average bubble area 
 $ \bar{A}^{-1/2}= 
1.7\cdot 10^4$ m$^{-1}$, and this theoretical expression would give a prediction,
$\mu = 37.2 $~nN$\cdot$m$^{-1}$, similar (within errors) to our measurement from 
image analysis. 
A previous experimental measurement on a dry (fluid fraction $<3$\%) 2D soap froth found $\mu$ higher than for the honeycomb of 20\%, and suggested to interpret it as an effect of disorder in wall lengths  \cite{asipauskas}. Here, we cannot discuss such effect, since it would be counterbalanced by the   effects of the fluid fraction (our 2D wet foam is expected to have a lower $\mu$ than the theoretical dry one \cite{weaire_bible}, similarly to  what happens in 3D \cite{durian,weitz}).

%\subsection{lien image-meca}

This local measurement of the shear modulus leads to a testable prediction: 
the value of the global foam rigidity, experienced by the free end 
of the fiber.  
The
foam rigidity
$k_{\mathrm{eff}}$ is a function of $\mu$ and $\nu$; due to the logarithmic range of 2D elasticity \cite{amorphes}, the rigidity also depends on the set-up
geometry  
%(Fig.  \ref{Fig set_up}) 
and   the boundary conditions at the edges of the trough, which we do not know. 
We use for simplicity the calculations for  non-slip boundary conditions 
\cite{amorphes}. Here 
the fiber radius is $r=3$~$\mu$m, the trough width  and length are
 $a = b =$ 2 cm, and we obtain:
\begin{equation}
k_{\mathrm{eff}} = \frac{\mu}{1.085 - 0.362 \; \nu}  .
\label{shear}
\end{equation}
The foam Poisson ratio  $\nu$, which is between $-1$ and 1,
is not measurable in  our experiment. Since it is probably determined 
mostly by the network of bubble walls (in the liquid-gas coexistence region, the pressure 
inside the bubbles is constant \cite{stine},
hence does not contribute to the foam compressibility), it is likely to be much smaller than 
unity \cite{boal}: this is compatible with the visual impression \cite{film_web}. If we 
neglected it for 
simplicity, eq. (\ref{shear}) would yield, using our measured value $\mu= 38 \pm 3\; 
\mathrm{nN}\cdot \mathrm{m}^{-1}$:
$$k_{\mathrm{eff}}  (\mathrm{image}) = 35 \pm 3\; \mathrm  {nN}\cdot \mathrm  {m}^{-
1}.$$

%\subsection{courbe meca}

This value predicted  under all the above assumptions can now be compared to the 
direct mechanical measurement of $k_{\mathrm{eff}}$, shown on
Fig.  (\ref{Fig_courbes}b).  
The subphase contribution to the force is negligible, 
as shown when performing the same experiment on the surface 
of pure water.
During the deformation  (Fig.  \ref{Fig_images}),
the foam apparently remains in its elastic regime. In fact, first, 
the  force-displacement plot is affine over a displacement of 3 mm,
 despite the  (yet unexplained) initial time-lag
visible on Fig.  (\ref{Fig_courbes}b). Second, 
as is visible on the movie  \cite{film_web},
no bubble rearrangement and no wall 
breakage is observed.
Third, even along the path of  the fiber,
the foam is intact:
when the fiber is displaced backwards,
the   force-displacement plot  is reversible.  
We find
that its  slope $k_{\mathrm{eff}} $
agrees (within errors) with the above prediction:
$$k_{\mathrm{eff}}  (\mathrm{force}) = 37.5 \pm 0.8\; \mathrm {nN}\cdot \mathrm 
{m}^{-1}.$$

%\section{Summary}
%
%%\subsection{message de l'auteur}
%
%
%In this letter, we simultaneously deform, image, and measure 
%mechanically the effective rigidity of an  intrinsically 2D foam,  in a Langmuir monolayer.  
%The deflection of a soft elastic fiber allows to 
%measure forces in the pN range, and to deduce precisely  the value 
%of the foam effective rigidity.
%
%Independently, using  images of the foam under deformation, we measure the gas-liquid 
%line tension  and then the stress tensor, as well as a new tool: the 
%statistical strain tensor  
%\cite{aubouy}. 
%Using a  method developed for soap froths \cite{asipauskas}, we perform an original 
%measurement  of 2D  shear elasticity and deduce the  shear 
%modulus  of the foam. Whether it depends on the bubble area distribution 
%(or only on the average area) is still an open question.
%
%From this value of $\mu$, {\it i.e.} from a local estimate based on image analysis,  we predict  
%the foam rigidity, which agrees with the direct mechanical 
%measurement.  

\section{Perspectives}

This result illustrates how  adapted is  the statistical strain 
tensor 
to describe locally the elastic properties 
of disordered  media. Beside foams, we intend to adapt it to other
amorphous materials, such as glasses and polymer networks, to investigate how the microscopic disorder of the structure affects the material's elasticity tensor.

\bigskip

%{\bf Acknowledgments}
\acknowledgments

We gratefully acknowledge the help and explanations of T. Fischer  for the measurement of 
the line tension.  
We thank  S. Akamatsu, M. Aubouy,
P. Ballet, S.  H\'enon, K. Kassner  and C. Quilliet for useful discussions.  This work was 
partially supported by CNRS ATIP 0693.


\begin{thebibliography}{0}
\bibitem{log_matrice}
By 
definition, the tensor  $\ln \overline{\overline{M}}$ is calculated by 
rotating $\overline{\overline{M}}$ to a basis where it is diagonal, 
taking the logarithm of its eigenvalues, then rotating it back into 
the initial basis.
\end{thebibliography}
\end{document}